# Theoretical study of hydrogen-covered diamond (100) surfaces: A chemical potential analysis

Suklyun Hong* and M. Y. Chou*

*School of Physics, Georgia Institute of Technology, Georgia 30332-0430*

(June 28, 1996)

The bare and hydrogen-covered diamond (100) surfaces were investigated through pseudopotential density-functional calculations within the local-density approximation. Different coverages, ranging from one to two, were considered. These corresponded to different structures including 1×1, 2×1, and 3×1, and different hydrogen-carbon arrangements including monohydride, dihydride, and configurations in between. Assuming the system was in equilibrium with a hydrogen reservoir, the formation energy of each phase was expressed as a function of hydrogen chemical potential. As the chemical potential increased, the stable phase successively changed from bare 2×1 to (2×1):H, to (3×1):1.33H, and finally to the canted (1×1):2H. Setting the chemical potential at the energy per hydrogen in $H_2$ and in a free atom gave the (3×1):1.33H and the canted (1×1):2H phase as the most stable one, respectively. However, after comparing with the formation energy of $CH_4$, only the (2×1):H and (3×1):1.33H phases were stable against spontaneous formation of $CH_4$. The former existed over a chemical potential range ten times larger than the latter, which may explain why the latter, despite of having a low energy, has not been observed so far. Finally, the vibrational energies of the C−H stretch mode were calculated for the (2×1):H phase.

## I. INTRODUCTION

Recently the nature of the structures of hydrogen on diamond (100) surfaces has attracted new interest due to technological advances of epitaxial growth of this surface by chemical vapor deposition (CVD) methods.[1−3] In the CVD homoepitaxial growth of diamond, the surface is mostly hydrogen terminated because abundant hydrogen exists in the growth environment.[2] Atomically smooth surfaces were obtained only in the (100) orientation.[4−6] Therefore, understanding the structural and electronic properties of this surface becomes quite important. Since there are two dangling bonds per surface carbon atom for the bulk-truncated bare surface, a reconstruction takes place in which two adjacent surface atoms move closer toward each other to form rows of symmetric dimers. Upon adsorption of hydrogen atoms, possible configurations include the monohydride (one hydrogen attached to a carbon atom) and dihydride (two hydrogen atoms attached to the same carbon atom) arrangements. The former is expected to keep the carbon dimer reconstruction (2×1), while the latter saturates both dangling bonds of a surface carbon atom, yielding a structure likely to be 1×1 and high in energy due to the steric repulsion.

Unlike the bare surface, what really happens on the hydrogen-covered C(100) surfaces was under debate for many years. A definite experimental determination of the stable structure was hindered by the difficulty of producing smooth surfaces, i.e. surfaces with large, atomically flat terraces.[7−10] Only recently have researchers resolved this problem.[4−6] Thomas and Butler[11] reported high resolution electron energy loss (HREEL) spectra indicating only monohydride species existed after *ex situ* exposure of the surface at 800°C to a hydrogen plasma and no dihydride was observed after *in situ* exposure of the surface to hydrogen atoms produced by a hot tungsten filament. In addition, they observed a sharp two-domain 2×1 reconstruction by low energy electron diffraction (LEED) for the hydrogen saturated surface. Küttel *et al.*[6] analyzed natural, doped (100) diamond surfaces polished in a hydrogen plasma by atomic force microscopy (AFM), LEED, and X-ray photoelectron diffraction. They found that after etching the C(100) surfaces had a roughness of 0.8 nm and showed a sharp 2×1 LEED pattern which was even stable in air without annealing, and that exposing these surfaces to atomic hydrogen produced by a heated filament led to an increase of surface roughness comparable to what was observed on the as-received sample. They concluded that the often reported 1×1 LEED pattern seemed to be a consequence of a large surface roughness and was absent on smooth surfaces.

Using the vibration sum-frequency generation (VSFG) spectroscopy, Anzai *et al.*[12] observed vibrational resonance peaks at 2910 and 2960 cm$^{-1}$. They assigned the main 2910 cm$^{-1}$ band to the CH stretch vibration of the monohydride groups on the C(100)(2×1) surface, and the 2960 cm$^{-1}$ band to the anti-symmetric CH stretching vibration of the dihydride groups located at either the unreconstructed C(100) (1×1) island surface or step sites. Kawarada *et al.*[13] examined the 2×1/1×2 surface reconstruction of C(001) surface using a scanning tunneling microscope (STM) at an atomic scale and reflection electron microscopy (REM) at a microscopic scale.

Quite a few theoretical studies of the structural and electronic properties of the H-covered C(100) surface have been performed,[14−23] including two recently published first-principles slab calculations[14,15] based on the local-density approximation (LDA). Furthmüller et al.[14] used the ultrasoft pseudopotentials[24] and the conjugate-gradient technique, while Zhang et al.[15] employed the Car-Parrinello molecular-dynamics (MD) method. Both examined the (2×1):H monohydride structure and found it to have a large desorption energy. Furthmüller et al.[14] also calculated a higher-coverage structure with 1.5 monolayer (ML, with respect to surface carbon atoms) of hydrogen on the surface, and found that the desorption energy for it to go to the (2×1):H monohydride phase was rather small. Zhang et al.[15] reported that the dihydride phase is energetically favored over the monohydride phase relative to H desorption, while it is unfavorable with respect to $H_2$ desorption. On the other hand, earlier studies by Yang et al.[16] using a nonself-consistent tight binding model with parameters obtained from density functional theory (DFT), and by Skokov et al.[17] using molecular dynamics with a semiempirical Hamiltonian, reported a (3×1):1.33H structure to be most favorable energetically. This structure consists of alternating monohydride carbon dimers and dihydride units. It was previously observed[25,26] in the hydrogen-covered Si(100) surface, but has not been identified experimentally for C(100). A fully self-consistent calculation is therefore necessary to clarify this situation. In addition, in order to compare stability of different phases with different hydrogen coverage, one has to take into account the chemical potential of hydrogen.[27] In this study, we will evaluate the energies of different structures with coverages ($\theta$) ranging from one to two ML, including all the phases mentioned above. Depending on the values of the chemical potential, the equilibrium phase in contact with a hydrogen reservoir can be the (2×1):H, (3×1):1.33H, or (1×1):2H (canted) phase. Comparing these formation energies with that of a stable hydrocarbon $CH_4$, one can exclude the (1×1):2H phase. Then the stable surface phase is (3×1):1.33H for a small chemical potential range and (2×1):H for a wide chemical potential range. We will also present the energetics in terms of the desorption energy and compare the results with previous studies. Finally, the vibrational energy of the C−H stretch mode is calculated for the (2×1):H phase.

## II. CALCULATIONS

The calculations were carried out within the local-density approximation (LDA)[28] with the Ceperley-Adler exchange-correlation potential[29] parametrized by Perdew and Zunger.[30] The soft pseudopotentials were generated by the scheme of Troullier and Martins[31] for both carbon and hydrogen. The carbon pseudopotentials were generated in the ground state configuration $2s^2 2p^2$. The cutoff radii were 1.50 and 1.54 a.u., respectively, for $s$ and $p$ potentials. It has been shown that good convergence can be achieved with a plane-wave cutoff of 49 Ry for these pseudopotentials.[31] The hydrogen pseudopotential was generated from the $1s^1$ configuration with a cutoff radius of 1.25 a.u. In the Troullier and Martin scheme, the first four derivatives of the pseudo-wave-function are matched to those of the all-electron wave function at the cutoff radius. Hence noticeable deviations of the pseudo-wave-function from the all-electron wave function normally occur at positions a few tenths of an a.u. inside the cutoff radii, and the cutoff radii chosen in this scheme are normally larger than those in other pseudopotential generation schemes. The wave functions in the surface calculation were expanded in plane waves with a cutoff $E_{cut} = 55$ Ry. To check if the cutoff energy is large enough for hydrogen, we calculated the total energy of a hydrogen molecule as a function of the cutoff energy. The difference in the total energy of $H_2$ between $E_{cut} = 55$ Ry and 100 Ry was below 10 meV per H. For diamond the lattice constant and the bulk modulus were found to be 3.53 Å and 4.56 Mbar, respectively, in good agreement with the experimental values of 3.567 Å and 4.43 Mbar. In the surface calculation, we used the theoretical lattice constant, which gave an ideal bond length of 1.53 Å.

Our slab model for the C(100) surfaces had the inversion symmetry and consisted of ten carbon layers and a vacuum region equal to six atomic layers (6.18 Å). We varied the thickness of the crystal slab and the vacuum region for the symmetric (1×1):2H phase, and found reasonable convergence of energy and geometry for our supercell. When varying the vacuum thickness, maximum changes were less than 0.02 eV/(1×1) in energy and about 0.5 % in bond length and angle using $E_{cut} = 55$ Ry. The atomic positions at the two innermost layers were kept fixed. For the Brillouin-zone integration we used 6×6×2, 3×6×2, and 2×6×2 grids in the Monkhorst-Pack special point scheme for the 1×1, 2×1, and 3×1 phases, respectively. For the metallic (2×1):1.5H phase, a 4×8×2 grid was also tested to get better convergence. We determined the hydrogen and relaxed carbon positions by minimizing the energy and Hellmann-Feynman forces. In our calculations, we used the force convergence criterion of 0.01 Ry/a.u., which gave a reasonable convergence in energy and geometry. Using a smaller force convergence criterion of 0.005 Ry/a.u. in the calculation of the symmetric (1×1):2H phase gave only a maximal change of 0.5% in the final bond length and an energy difference of 0.02 eV per (1×1) surface cell.

For the bare surface and the (2×1):1.5H phase, we used Gaussian broadening with a width of 0.1 eV to accelerate the convergence in the $k$-point sum, because these structures have dangling bonds on C atoms in intermediate or final configurations. We used a precondition Lanczos subspace-diagonalization procedure to solve for the eigenvalues and eigenvectors of the Hamiltonian matrix for a given LDA potential, and a modified quasi-Newton scheme to find the self-consistent potential.[32]

## III. RESULTS AND DISCUSSIONS

### A. Structures

We have studied structures of different coverages and atomic configurations for the hydrogen-covered C(100) surfaces, including: (a) the bare reconstructed 2×1 surface, (b) the (2×1):H surface, (c) the (2×1):1.5H surface, (d) the (3×1):1.33H surface with alternating monohydride dimer and dihydride units, and (e) the symmetric, (f) canted, and (g) rotated (1×1):2H surfaces. We note that the unsaturated (2×1):1.5H surface was identical to the (1×1):1.5H surface considered by Furthmüller et al.[14]. We used the (2×1) notation to represent the periodicity of the surface unit cell. The top and side views of these structures are shown in Fig. 1.

The calculated bond lengths for the top layers are given in Table I and compared with previous calculations. Our C−H and top-layer C−C bond-length results were almost identical to those of earlier LDA calculations of Furthmüller et al.[14] (using ultrasoft pseudopotentials and the LDA) and Zhang et al.[15] (using ab initio MD) for structures they had considered.

For the bare 2×1 surface, we obtained the symmetric dimer as the ground state. We started with a buckled structure and the system relaxed to the symmetric one. This symmetric dimer structure is in contrast to that of Si(100) with a c(4×2) reconstruction composed of asymmetric dimers.[33] The origin of asymmetry has been identified as a result of competition between the $\pi$ bond and Jahn-Teller-like distortion.[34,35] For the bare C(100)(2×1), the filled $\pi$-bonding states were found to be separated from the empty antibonding $\pi^*$ states even for symmetric dimers. Therefore, no Jahn-Teller type distortion occurred. Similar to previous findings, the C−C dimer bond length on the bare C(100)(2×1) surface (1.37 Å) was close to that of a typical carbon double bond. This strong $\pi$ and $\sigma$ bonding disappeared upon hydrogen adsorption in the (2×1):H phase in which the C−H bond was formed. The C−C dimer bond length significantly lengthened to 1.63 Å, being even larger than the bulk C−C bond length.

For the (3×1):1.33H phase the bond lengths of the monohydride and dihydride were very similar to corresponding ones in the (2×1):H and symmetric (1×1):2H phases, respectively, as shown in Table I. This structure is expected to have a low energy because the large steric repulsion caused by hydrogens in the (1×1):2H phase is greatly reduced by the presence of the dimers.[16] The energetics will be considered in the next section. The (2×1):1.5H phase was considered by Furthmüller et al.[14] as a possible high-coverage phase. We obtained almost identical geometries compared with their results. Since unsaturated dangling bonds exist in this structure, it is not energetically favorable as will discussed in the next section.

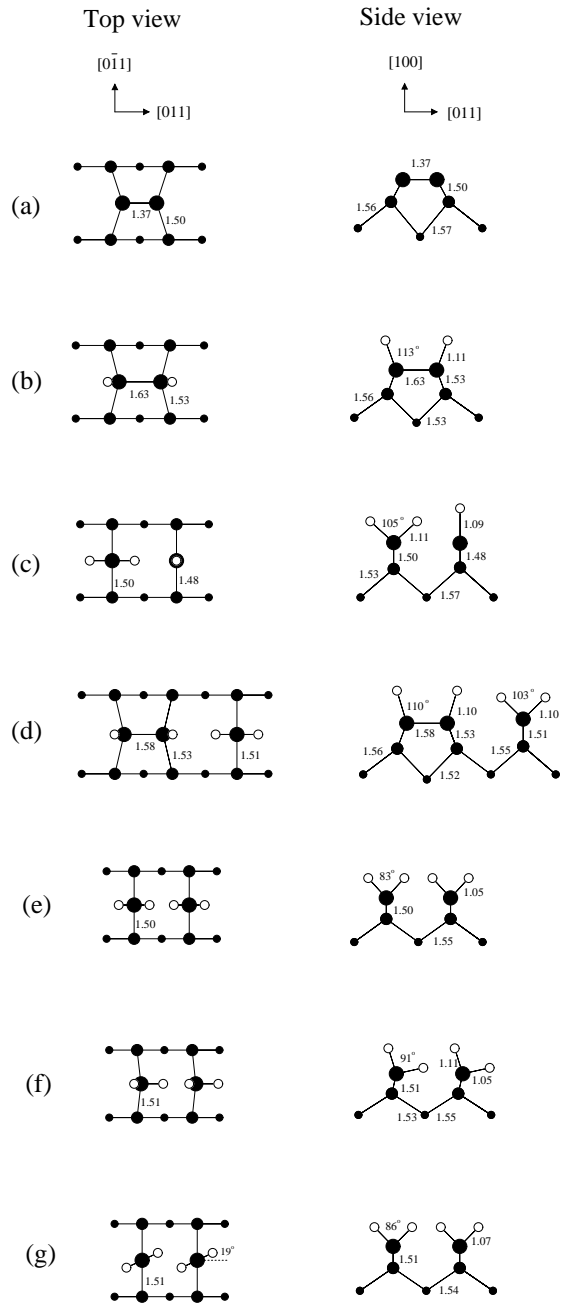

FIG. 1. Top and side views of C(100) surfaces with different hydrogen configurations: (a) bare 2×1, (b) (2×1):H, (c) (2×1):1.5H, (d) (3×1):1.33H, (e) symmetric (1×1):2H, (f) canted (1×1):2H, (g) rotated (1×1):2H. The bond lengths, in units of Å, and bond angles are shown. Comparisons with previous theoretical results are given in Tables I and II.

For completeness we also considered the structures of the (1×1):2H surfaces. A symmetric configuration [Fig. 1 (e)] has a large steric repulsion between two hydrogens attached to adjacent carbon atoms. This repulsion can be reduced by a canted-row structure obtained from the symmetric dihydride by rotating the CH$_2$ groups around an axis passing through the second layer carbon atoms [Fig. 1 (f)]. A rotation of ∼10° increases the distance between hydrogen atoms in the same (neighboring) dihydrides from 1.40 to 1.55 Å (from 1.10 to 1.44 Å), thereby lowering the energy. This canted dihydride structure was found to be 0.42 eV/(1×1) lower in energy than the symmetric dihydride. Alternatively the steric repulsion can be reduced by rotating the dihydride units around an axis perpendicular to the surface [Fig. 1 (g)], but we found that the energy gain was quite small, only 0.01 eV per 1×1 unit cell. Also, we considered an alternating 1×2 canted structure, which turned out to be 0.07 eV/(1×1) higher in energy than the original 1×1 canted one.

TABLE I. Calculated bond lengths (in Å) in top two layers for various hydrogen-covered C(100) surfaces compared with results from other density-functional calculations. More calculated bond lengths are shown in Fig. 1. For comparisons with other previous theoretical studies, see Ref. 14.

|  | This work | Ref. 14[a] | Ref. 15[b] | Ref. 16[c] |
|---|---|---|---|---|
| Lattice constant | 3.53 | 3.53 | 3.56 | 3.62 |
| Bare 2×1 | symmetric | symmetric | symmetric | buckled |
| $d_{11}$(C−C) | 1.37 | 1.37 | 1.38 | 1.40 |
| $d_{12}$(C−C) | 1.50 | 1.50 | − | 1.52, 1.55 |
| (2×1):H |  |  |  |  |
| $d$(C−H) | 1.11 | 1.10 | 1.11 | 1.17 |
| $d_{11}$(C−C) dimer | 1.63 | 1.61 | 1.63 | 1.67 |
| $d_{12}$(C−C) | 1.53 | 1.53 | − | 1.59 |
| (3×1):1.33H |  |  |  |  |
| $d$(C−H) monohydride | 1.10 | − | − | 1.18 |
| $d$(C−H) dihydride | 1.10 | − | − | 1.17 |
| $d_{11}$(C−C) dimer | 1.58 | − | − | 1.62 |
| $d_{12}$(C−C) monohydride | 1.53 | − | − | 1.58 |
| $d_{12}$(C−C) dihydride | 1.51 | − | − | 1.60 |
| (2×1):1.5H |  |  |  |  |
| $d$(C−H) monohydride | 1.09 | 1.09 | − | − |
| $d$(C−H) dihydride | 1.11 | 1.11 | − | − |
| $d_{12}$(C−C) monohydride | 1.48 | 1.47 | − | − |
| $d_{12}$(C−C) dihydride | 1.50 | 1.50 | − | − |
| (1×1):2H (symmetric) |  |  |  |  |
| $d$(C−H) | 1.05 | − | 1.06 | − |
| $d_{12}$(C−C) | 1.50 | − | − | − |
| (1×1):2H (canted) |  |  |  |  |
| $d$(C−H) | 1.11, 1.05 | − | 1.04−1.13 | − |
| $d_{12}$(C−C) | 1.51 | − | − | − |

[a] Furthmüller et al., self-consistent DFT
[b] Zhang et al., ab initio MD-DFT
[c] Yang et al., non-self-consistent DFT

Since different lattice constants were used in different theoretical calculations, a comparison of angles may be helpful. The angles H−C−C and H−C−H are given in units of degrees in Table II and shown in Fig. 1, along with the canting angle in the canted (1×1):2H phase. Our results are also similar to those of Furthmüller et al.[14] and Zhang et al.[15] for the structures they had considered.

TABLE II. Calculated angles in units of degrees (shown in Fig. 1) in various hydrogen-covered C(100) surfaces compared with results from some previous theoretical studies. The H−C−H angle for the canted dihydride phase of Ref. 15 is for an alternating canted and twisted 1×2 structure.

|  | This work | Ref. 14 | Ref. 15 | Ref. 16 |
|---|---|---|---|---|
| (2×1):H |  |  |  |  |
| H−C−C | 113 | 113.3 | 113.1 | 113.9 |
| (3×1):1.33H |  |  |  |  |
| H−C−H dihydride | 103 | − | − | 108.7 |
| H−C−C monohydride | 110 | − | − | 108.7 |
| (2×1):1.5H |  |  |  |  |
| H−C−H | 105 | 103.1 | − | − |
| (1×1):2H |  |  |  |  |
| H−C−H (symmetric) | 83 | − | 84.72 | − |
| H−C−H (canted) | 91 | − | 93−96 | − |
| Canting angle | 11 | − | − | − |

## B. Energetics

Most previous calculations agreed that the dihydride structure had a high energy because of the steric repulsion between two nearby hydrogen atoms, and that the (2×1):H monohydride structure, with all dangling bonds saturated, was a reasonably stable phase. Some calculations also considered other competitive structures with higher hydrogen coverages. Furthmüller et al.[14] studied the (2×1):1.5H phase. Yang et al.[16] using a nonself-consistent approach and Skokov et al.[17] using a semiempirical Hamiltonian reported that the (3×1):1.33H structure with alternating monohydride dimers and dihydride units was quite low in energy. Zheng and Smith[18] found a strained symmetric dihydride configuration as stable using slab-MINDO (modified intermediate neglect of differential overlap). In order to compare stability among different phases with different coverages, we will first adopt a chemical potential analysis[27] by considering the system in equilibrium with a hydrogen reservoir characterized by a chemical potential $\mu_H$. Later, the desorption energy results will be discussed.

The formation energy $\Omega$ was defined as[27]

$$\Omega = E + E_0 - n_C \mu_C - n_H \mu_H, \quad (1)$$

where $E$ was the total energy at zero temperature, $E_0$ the zero-point energy, and $n_C$ and $n_H$ the numbers of carbon and hydrogen atoms, respectively. Since the systems under study were assumed to be in chemical and thermal equilibrium with bulk carbon, we used the bulk energy as the chemical potential of carbon, $\mu_C$. The zero-point energy included was from C–H vibrational modes. Since the hydrogen is always bonded to a single C atom, so it is reasonable to assume a zero-point energy proportional to the number of H atoms:[27] $E_0 = n_H e_0$. We estimated $e_0$ by the sum of vibrational frequencies of a typical hydrocarbon $CH_4$ and set it to 0.293 eV (0.0215 Ry) per hydrogen.[36] Temperature dependence of the formation energy was mainly incorporated through the chemical potential of hydrogen. Although CVD is a highly nonequilibrium process, the study of equilibrium systems would be the first step in understanding stability of different phases during the growth.

The calculated formation energies per 1×1 unit cell for different phase are shown in Fig. 2 as a function of the H chemical potential. The horizontal line corresponds to the bare 2×1 surface and the formation energy takes the value of the surface energy per 1×1 unit cell. Our result of 2.18 eV was close to the value of 2.12 eV obtained in Ref. 14. From Fig. 2, we can see that stability of each phase depends on the chemical potential of hydrogen. As the H chemical potential increases, the bare 2×1, (2×1):H, (3×1):1.33H, and the canted dihydride phase successively become the most stable phase. The phase boundaries, indicated by arrows in Fig. 2, are located at −17.19 eV, −15.66 eV, and −13.54 eV, respectively. This behavior is similar to that in the H-covered

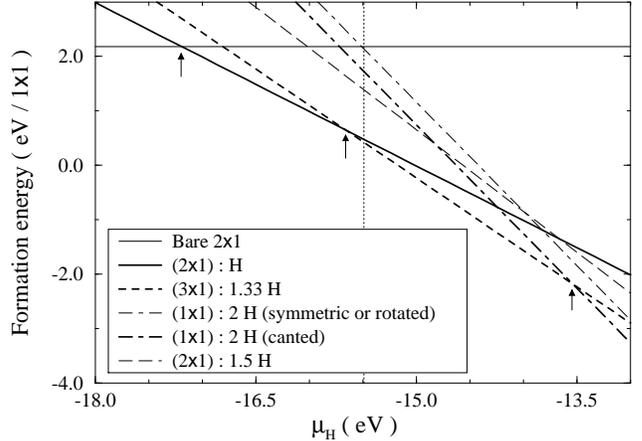

FIG. 2. Formation energies $\Omega$ [Eq. (1)], in units of eV per 1×1 cell, for various phases as a function of hydrogen chemical potential $\mu_H$. As $\mu_H$ increases, the bare 2×1, (2×1):H, (3×1):1.33H, and canted (1×1) phases successively become the phase with the the lowest formation energy. Arrows denote the phase boundaries at −17.19 eV, −15.66 eV, and −13.54 eV. The vertical dotted line at −15.49 eV, just beside the cross point of the (2×1):H and (3×1):1.33H phases, indicates the chemical potential which gives a zero formation energy for $CH_4$. See text for details.

Si(100) surfaces reported in Ref. 27. The formation energies for two selected chemical potential values, −0.957 Ry (−13.02 eV) and −1.127 Ry (−15.33 eV), are listed in Table III, along with the total energies for each phase. The former is the energy of a hydrogen pseudoatom calculated with the local spin-density approximation (LSD), while the latter is the calculated energy per hydrogen in $H_2$ with the zero-point energy of the vibrational mode (4400 $cm^{-1}$)[37] included. In the first case, the canted (1×1):2H is most stable, while the (3×1):1.33H has the lowest formation energy in the second case.

Not all the surface phases can be realized after taking into account other competing processes on the surface, such as the formation of hydrocarbons, within the same framework. We calculated the formation energy of $CH_4$ with the same theoretical methods using a supercell. The vertical dotted line in Fig. 2 near the crossing point of the (2×1):H and (3×1):1.33H phases represents a hydrogen chemical potential at which $CH_4$ molecules can be formed with no cost in energy. The value is $[E(CH_4) + 4e_0 - \mu_C]/4 = -15.49$ eV. On the right hand side of the vertical line, the formation energy of $CH_4$ is negative, namely, $CH_4$ forms spontaneously from the C(100) surfaces in the presence of the hydrogen reservoir. In contrast, the left side of the vertical line has a positive formation energy for $CH_4$. The most stable surface phases for this region are: the bare surface when $\mu_H$ is smaller than −17.19 eV, the (2×1):H phase when $-17.19 < \mu_H < -15.66$ eV, and a small window for the (3×1):1.33H phase when $-15.66 < \mu_H < -15.49$ eV (see Fig. 2). The dihydride phase region is totally excluded, as well as most of the 3×1 phase. Thus for a large $\mu_H$ range (about 1.5 eV) the (2×1):H phase has the lowest formation energy, while the range for the (3×1):1.33H phase is quite limited (about 0.17 eV). In comparison, for the hydrogen covered Si(100) surfaces the hydrogen chemical potential that gives zero formation energy for SiH$_4$ lies within a region where the dihydride is most stable.[27]

TABLE III. Total energy $E$ (in Ry) and formation energy $\Omega$ (in eV, for two selected values of the hydrogen chemical potentials) per (1×1) unit cell for various C(100) surfaces.

|  | Bare 2×1 | (2×1):H | (3×1):1.33H | (2×1):1.5H | (1×1):2H | |
|---|---|---|---|---|---|---|
|  |  |  |  |  | symmetric | canted |
| $E$ | −57.029 | −58.314 | −58.705 | −58.824 | −59.352 | −59.383 |
| $\Omega^a$ | 2.18 | −2.00 | −2.88 | −2.28 | −2.81 | −3.23 |
| $\Omega^b$ | 2.18 | 0.31 | 0.20 | 1.19 | 1.81 | 1.40 |

[a] using $\mu_H = -13.02$ eV, the LSD value for a free hydrogen atom
[b] using $\mu_H = \frac{1}{2}E(H_2) = -15.33$ eV

TABLE IV. Desorption energy, in units of eV, of a hydrogen atom (see text) for various hydrogen-covered surfaces. The value for the (2×1):H phase is in reference to the bare (2×1) surface, while the values for other phases are with respect to the (2×1):H phase. The listed values were obtained using the LSD atomic energy of −13.02 eV, and did not include the zero-point energy of C−H vibrations.

|  | (2×1):H | (3×1):1.33H | (2×1):1.5H | (1×1):2H | |
|---|---|---|---|---|---|
|  |  |  |  | symmetric | canted |
| This work | 4.47 | 2.93 | 0.86 | 1.10 | 1.52 |
| Ref. 14 | 4.54 | − | 1.02 | − | − |
| Ref. 15 | 5.49 | − | − | 1.13 | − |
| Ref. 16 | 6.20 | 3.90 | − | − | − |

In general, the formation energy of defects involving higher hydrides will be reduced as $\mu_H$ increases. Therefore, the (1×1):2H and (3×1):1.33H phases, which would become stable at larger $\mu_H$, are expected to be more susceptible to defect formation than the (2×1):H phase. These defects increase disorder on the surface, and may yield a 1×1 LEED pattern due to the underlying lattice as viewed through the disordered top layer. The relatively small $\mu_H$ region for a stable 3×1 phase and the consideration of defect formation might explain why the C(100)(3×1):1.33H has not been observed so far.[38]

To answer the question whether other n×1 phases can be realized, we estimated their energies by a simple model used in Ref. 27. It was based on the assumption that the formation energy could be approximated as a sum of the formation energies of isolated structural subunits and the interaction energy between these subunits. Define $U$ as the difference in formation energy between the canted 1×1 phase and an isolated dihydride row unit. It is the energy arising from interactions between neighboring rows of dihydrides, including the relaxation energy due to the CH$_2$ rotation. We estimated that $U = 1.41$ eV, compared with that of 0.10 eV for Si(100).[27] This large repulsive interaction resulted in the wide range of the 3×1 phase. Using this model, the stability of any n×1 structure consisting of a mixture of monohydride and dihydride units can be estimated. One type of them ($n \geq 4$) contains one monohydride row and $n-2$ canted dihydride rows in each cell, and another consists of one dihydride row and $(n-1)/2$ monohydride rows. It can be easily shown that both of these n×1 structures for C(100) surfaces are unstable relative to the phases we considered, same as concluded in Ref. 27 for Si(100).

Another way to examine the energetics is to calculate the desorption energy for these systems. Following previous theoretical studies, the desorption (adsorption) energy $\Delta E_H$ per hydrogen atom from the (2×1):H surface is defined as the energy difference per surface site between the hydrogenated, reconstructed surface and the bare, reconstructed surface plus a free hydrogen atom We used the LSD energy of −13.02 eV for a free hydrogen atom. The desorption energies per hydrogen atom for higher-coverage structures were calculated with reference to the (2×1):H phase, instead of the bare surface. They can also be obtained by taking the difference of corresponding formation energies listed in Table III, and correcting for the zero-point energy and differences in coverage. Table IV summarizes these results and compares them with those obtained from other density-functional calculations. More realistic desorption energy values should include the zero-point energy of C−H vibrational modes (about 0.29 eV per hydrogen) and reference to the energy of H$_2$. Those values can be obtained by shifting down our results in Table IV by a constant of 2.60 eV.

From Table IV, we see that our calculated desorption energies are close to those of Furthüller et al.[14] for systems both considered. A larger difference was found between our (2×1):H result and that of Zhang et al.,[15] which may in part result from the Γ point sampling in Ref. 15. The difference between our results and those from the nonself-consistent calculation by Yang et al.[16] was even bigger.

TABLE V. Calculated vibrational energies of the H−C stretch mode in the (2×1):H phase. Results are obtained using the reduced-mass approximation and the masses of $^1$H and $^{12}$C. Energies are given in units of cm$^{-1}$, while the values in parentheses are in meV.

|  | Theory | | Experiment | | |
| --- | --- | --- | --- | --- | --- |
|  | This work | Ref. 40 | Ref. 11 | Ref. 12 | Ref. 10 |
| Ground to first | 2835 (351) | 2750 | 2915 | 2910 | 2928 |
| Ground to second | 5550 (688) | − | − | − | − |
| Anharmonicity | −120 (−15) | − | − | − | − |

### C. Vibrational frequencies for the monohydride surface

We have studied the hydrogen vibrational frequencies for the C(100)(2×1):H surface within the frozen-phonon approximation by considering the energy as a function of hydrogen displacement from its equilibrium position along the stretch direction, with the carbon atoms fixed. This is the ordered structure observed experimentally on smooth surfaces so far. Since the hydrogen potential could not be well fitted by a harmonic form at larger atomic displacement, we fitted it by a quartic polynomial

$$V(r) = V_0 + 1.124r^2 - 2.310r^3 + 2.292r^4 \quad (2)$$

where $V(r)$ is in Ry and the displacement $r$ is in Å. The plot is shown in Fig. 3. The Schrödinger equation of a hydrogen atom with this anharmonic potential was then solved numerically. The mass of the hydrogen atom must be reduced because of substate motion. In our calculation, it was replaced by the reduced mass of a hydrogen atom and a carbon atom. It has been shown[39] that the effective-mass approximation is sufficient for the H−C stretch mode on the hydrogen covered C(111) surface. Higher order corrections were estimated to be only 0.35 % in Ref. 39 by a classical model with the spring constant between the top-layer carbon atoms and the substrate calculated from first principles. We performed a similar calculation for the C(100) (2×1):H surface and found that the higher-order correction to be about 0.3 % within the simple classical model. The calculated energy separations between the ground and first excited states are shown in Table V, compared with other theoretical and experimental values. Our result was about 3 % smaller than the experimental values, similar to the difference between the LDA result and experiment for hydrogen on C(111).[39] The local-density approximation may be responsible for this difference, or the reduced-mass approximation may have a larger error than estimated by the simple classical model.[39] The calculated H−C stretch mode frequency is larger on C(100) than on C(111) by about 3 %, same as the ratio of measured values for the two surfaces, although the calculated bond-length difference on these two surfaces is no more than 1 %. We also examined the anharmonicity of the stretch mode, which is usually measured by the difference between the excitation energy from the ground state to the second excited state and twice the energy from the ground sate to the first excited state. The result, shown in Table V, was about 2 %, again similar to what was found for hydrogen on C(111).[39]

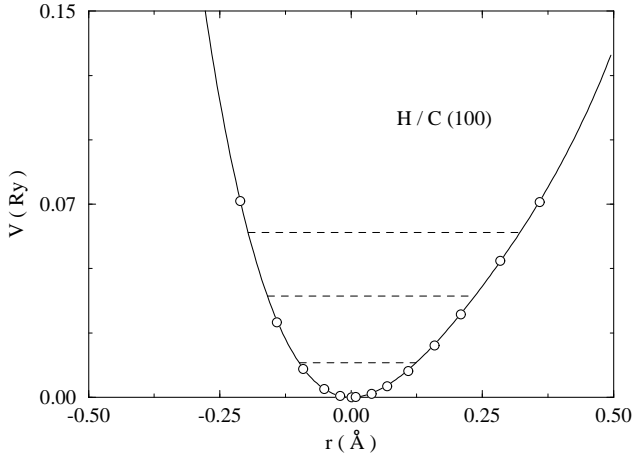

FIG. 3. Calculated energy for the H−C bond stretch mode in the C(100)(2×1):H phase as a function of H displacement $r$ (in units of Å) from its equilibrium position along the stretch direction. The solid line is a fourth-order polynomial fit, while the open circles represent calculated energies. The first three levels of vibrational energies are shown as horizontal lines.

### IV. SUMMARY

We have studied the bare and hydrogen-covered C(100) surfaces using pseudopotentials within the LDA frame-work. Different hydrogen coverages from one to two ML were considered. First, total energies and geometries of each phase were calculated. Relative stability between different phases with different coverages was investigated by examining the formation energy as a function of the hydrogen chemical potential. Taking into account the formation energy of $CH_4$, two stable phases, (2×1):H and (3×1):1.33H, were found in different chemical potential ranges. The former was found in a wider range of $-17.19 < \mu_H < -15.66$ eV, while the latter in a relatively small region of $-15.66 < \mu_H < -15.49$ eV. The region in which the (3×1):1.33H has the lowest formation energy was originally much larger, but was significantly reduced when the formation energy of $CH_4$ was taken into account. This small region coupled with a higher possibility of defect formation might explain why the 3×1 phase has not been observed so far. Finally, we calculated the vibrational energies of C−H stretch mode for the (2×1):H phase, and found them to be slightly larger than those on the C(111) surface.

## ACKNOWLEDGMENTS

We thank Dr. John Northrup for helpful discussions. This work was supported in part by the NSF (DMR-9157537) and DOE (DE-FG05-90ER45431). M. Y. C. thanks the Packard Foundation for its support.

[*] Electronic address: ph279sh@cmt6.physics.gatech.edu (S.H.); ph279mc@prism.gatech.edu (M. Y. C.)